\documentclass[twocolumn]{aastex631}
\usepackage{amsmath}
\usepackage{amssymb}
\usepackage{mathrsfs}
\usepackage{xcolor}
\usepackage{hyperref}
\usepackage{physics}
\usepackage{simplewick}
\usepackage{epsfig}
\usepackage{array}
\usepackage{booktabs}
\usepackage{subfigure}
\usepackage{gensymb}

\newcommand{\R}[1]{\textcolor{red}{#1}}
\colorlet{DeepBlue}{blue!80!black} 
\newcommand{\B}[1]{\textcolor{DeepBlue}{#1}}

\begin{document}

\title{Simulations of Tidal Disruption of Supernova in Galaxy Nuclear Region: A Novel Model for Ambiguous Nuclear Transients}

\author[0009-0003-0516-5074]{Xiangli Lei}
\affiliation{Department of Astronomy, School of Physics, Huazhong University of Science and Technology, Luoyu Road 1037, Wuhan, China}

\author[0000-0003-4773-4987]{Qingwen Wu$^*$}
\email{* Corresponding author: qwwu@hust.edu.cn} 
\affiliation{Department of Astronomy, School of Physics, Huazhong University of Science and Technology, Luoyu Road 1037, Wuhan, China}

\author[0000-0002-7329-9344]{Ya-Ping Li}
\affiliation{Shanghai Astronomical Observatory, Chinese Academy of Sciences, Shanghai 200030, People’s Republic of China}

\author[0000-0003-3440-1526]{Wei-Hua Lei}
\affiliation{Department of Astronomy, School of Physics, Huazhong University of Science and Technology, Luoyu Road 1037, Wuhan, China}

\begin{abstract}

An increasing number of ambiguous nuclear transients, including some extreme nuclear transients with very shallow light-curve declines and weak AGN activity in their host galaxies, have been reported. Stars form in or are captured by AGN disks will grow and migrate inward, potentially exploding as supernovae once the inner cold accretion disk disappears in low-luminosity AGNs. We propose that the tidal disruption of a supernova (TDS) by a supermassive black hole (SMBH) can produce nuclear transients that are more energetic and evolve more slowly than typical tidal disruption events (TDEs), without the black hole mass limit as in TDEs. In this scenario, the SMBH capture the supernova ejecta, which subsequently self-intersects and circularizes into an accretion disk. Based on hydrodynamical simulations, we find that the accretion rate of the TDS disk exhibits a slow decline that can last for months to decades. The peak accretion rate of a typical core-collapse SN scenario can exceed the Eddington limit for SMBHs with $M_{\rm BH} \lesssim 10^{7.5}\,M_\odot$, while it remains sub-Eddington for more massive SMBHs. This model provides a mechanism for triggering an energetic TDE-like flare with luminosity \(\gtrsim10^{45}\,\mathrm{erg\,s^{-1}}\) in weak AGNs even with SMBH mass much larger than $10^{8}\,M_\odot$ or triggering turn-on changing-look AGNs.

\end{abstract}

\keywords{Accretion (14), Active galactic nuclei (16), Supernovae(1668), Tidal disruption (1696), Hydrodynamics (1963)}

\section{Introduction} \label{sec:intro}

\begin{figure} 
  \centering
  \includegraphics[width=1.0\linewidth]{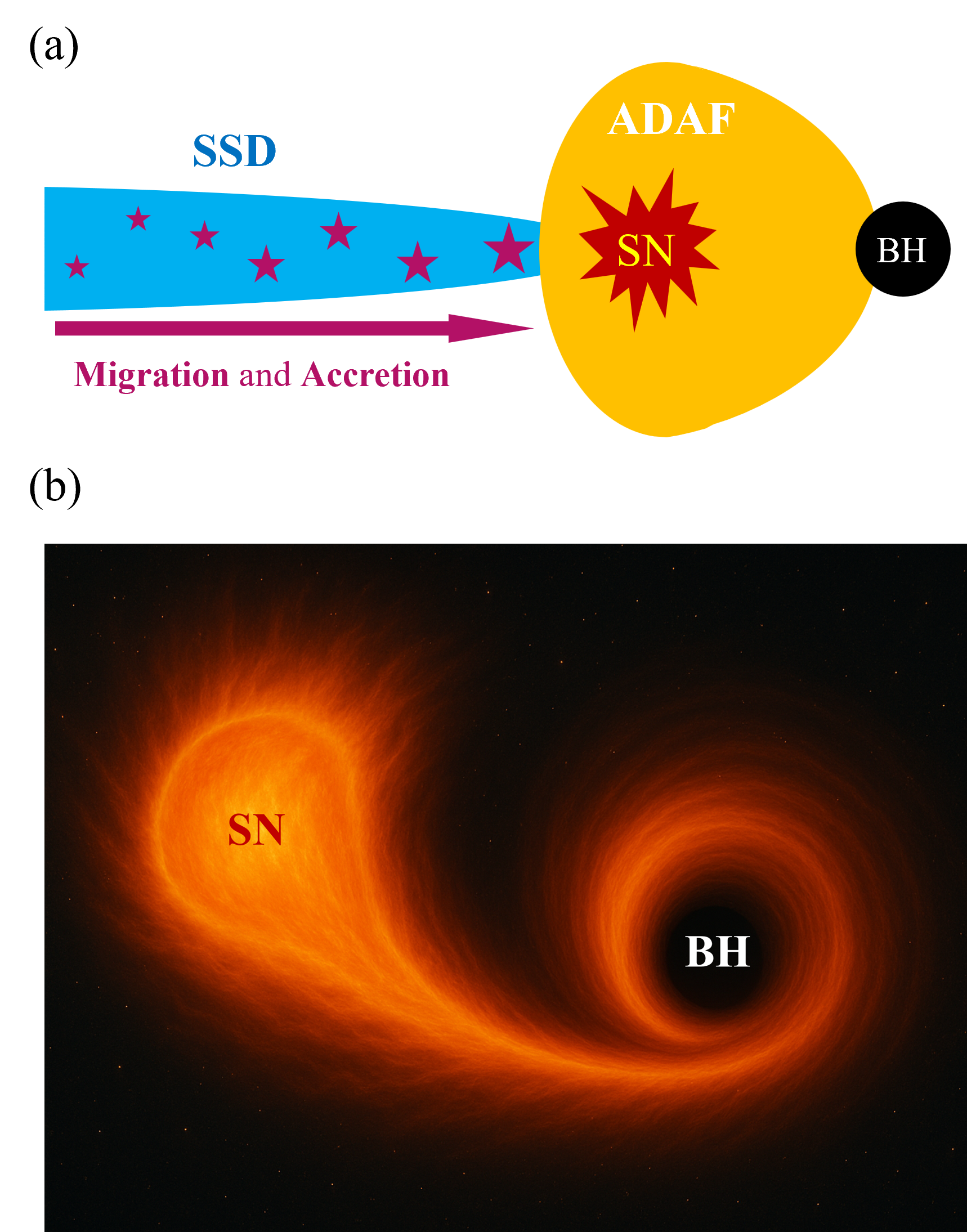}
  \caption{Schematic diagram of tidal disruption of a supernova (TDS). (a) Stars embedded in AGN disk accrete and grow during their inward migration. The massive disk stars will roughly stop accretion, quickly evolve and eventually explode as SN after the star migrate into inner low-density ADAF region. (b) After the SN explosion, part of the ejecta is captured by the central supermassive BH. The resulting bound streams undergo self-intersection and eventually circularize to form an accretion disk.}
  \label{fig:cartoon}
\end{figure}

Accretion onto supermassive black holes (SMBHs) powers many of the most luminous events in the universe. Continuously accreting SMBHs are observed as active galactic nuclei (AGNs) that present in $\sim$1-5\% of galaxies in the local universe \citep[e.g.,][]{Ho2008,Haggard2010}. In the era of multi-wavelength transient surveys, an increasing number of flares originating from galactic nuclei have been detected \citep[e.g.,][]{York2000,Chambers2016,Kochanek2017,Bellm2019,Yuan2022}. The tidal disruption of stars by SMBHs, known as tidal disruption events (TDEs) \citep[e.g.,][]{Rees1988,Evans1989,Gezari2021}, has been extensively studied in recent years, with more than two hundred cases reported \citep[e.g.,][]{Komossa2015,Hammerstein2023,Yao2023,Langis2025}. Apart from TDEs, some ambiguous nuclear transients (ANTs) or extreme nuclear transients (ENTs) have been discovered \citep[e.g.,][]{Neustadt2020,Hinkle2022,Subrayan2023,Wiseman2025,Hinkle2025}, which appear to be related to SMBHs but display observational properties distinct from both TDEs and AGNs. For example, a subset of TDE-like flares occurs in galaxies that host SMBHs with $M_{\rm BH}\gtrsim10^{8}\,M_\odot$ and shows peak luminosities $\gtrsim10^{45}\,\rm erg\,s^{-1}$ \citep[e.g.,][]{Leloudas2016,Hinkle2025,Graham2025}. In these cases, the luminosities are significantly higher than typical TDEs, and the SMBHs are too massive to tidally disrupt a sun-like star. Moreover, many extremely variable quasars (EVQs) and changing-look AGNs (CLAGNs) have been identified in optical and X-ray surveys, exhibiting variability distinct from the stochastic fluctuations typical of AGNs \citep[e.g.,][]{MacLeod2016,Rumbaugh2018,Trakhtenbrot2019,Ricci2023,Li2024}. The physical mechanisms that drive ANTs, ENTs, CLAGNs, and EVQs remain very unclear.

Galactic nuclei commonly host compact, massive nuclear star clusters (NSCs) that are resolved in many nearby galaxies \citep[e.g.,][]{Schodel2007,Genzel2010}. The NSCs are also closely correlated with extreme gravitational phenomena (e.g., extreme mass-ratio inspirals, EMRIs) and TDEs (see \citealt{Neumayer2020}, for a review). In AGNs, gravitational instability in the outer regions of the accretion disk can promote in-situ star formation, and some stars above or below the disk can also be captured. The rapid accretion and evolution of disk stars can lead to rapid metal enrichment as suggested by the high metallicities observed in the broad-line region \citep[e.g.,][]{Paczynski1978,Goodman2004,Fan2023}. Such a dense disk environment further promotes stellar accretion and inward migration \citep[e.g.][]{Liy2024,Laune2024}. These massive stars can reach the inner region within a few hundred gravitational radii \citep[e.g.,][]{Bartko2010,Bellovary2016,Pan2021,Cantiello2021}. After the fade of AGNs, the inner cold Shakura-Sunyaev disk (SSD) will disappear or transit to an advection-dominated accretion flow (ADAF) at tens to hundreds of gravitational radii when the accretion rate falls below a critical value (e.g., $\sim 1\%$ Eddington accretion rate) \citep[e.g.,][]{Ho2008,Yuan2014}. In this case, the former disk stars will stop accretion and evolve quickly, and their lifetimes are short due to their high mass. In this scenario, these stars may explode as supernovae in the fading phase of AGNs. The environment near an SMBH may cause SN ejecta to evolve differently from an isolated explosion \citep[e.g.,][]{Burrows2021,Li2023,Fabj2025}. 

If a star explodes sufficiently close to the SMBH, the SMBH can capture a large fraction of the SN ejecta and tidally stretch it into an elongated stream (see schematic diagram in Figure~\ref{fig:cartoon}). We refer to it as the tidal disruption of a supernova (TDS). The resulting streams circularize and form an accretion disk that can power very luminous nuclear flares, potentially explaining some energetic ANTs or ENTs \citep[e.g.,][]{Leloudas2016,Holoien2022,Hinkle2025}. To explore this process, we perform three-dimensional hydrodynamic simulations of TDSs. Section~\ref{sec:method} describes the numerical methods, Section~\ref{sec:model} presents the results, and Section~\ref{sec:discussion} summarizes our conclusions and discusses the observational features and possible counterparts.

\section{Numerical Methods} \label{sec:method}

SN explosions are generally isotropic, but the tidal field of a nearby SMBH can strongly break this symmetry and alter the ejecta evolution. To investigate this effect, we perform three-dimensional hydrodynamic simulations of SN ejecta launched from the mid-plane of a disk around an SMBH. As a fiducial case (\textit{Model-0}), we consider a faded AGN hosting a central SMBH with \(M_{\rm BH}=10^8\,M_\odot\) and an Eddington ratio \(f_{\rm Edd}=\dot{M}/\dot{M}_{\rm Edd}=10^{-3}\). In this scenario, the inner region of the disk is expected to stay in ADAF state. We therefore adopt the ADAF gas distribution as the SN environment, using the self-similar ADAF solution with a dimensionless shear viscosity parameter \(\alpha = 0.2\) \citep{Narayan1994,Nelson2013}. For the SN ejecta, we adopt an isotropic exponential density profile with a linear velocity structure, commonly used in analytical and numerical models \citep[e.g.,][]{Nomoto1984,Matzner1999,Grishin2021}. In our simulations, we adopt an ejecta mass of $M_{\rm ej}=10\,M_\odot$ and a kinetic energy of $E_{\rm ej}=2\times10^{51}\,\mathrm{erg}$, corresponding to a characteristic expansion velocity of $v_{\rm ej}= \sqrt{E_{\rm ej}/M_{\rm ej}}\simeq3.2\times10^8\,\mathrm{cm\,s^{-1}}$ with \(E_{\rm ej, kinetic} = E_{\rm ej, thermal}\), which are typical values for core-collapse supernovae \citep[e.g.,][]{Burrows2021}. The explosion is initiated at a fiducial radius $R_{\rm ej}=100\,R_{\rm g}$, which roughly correspond to a location near the expected transition radius of the accretion disk \citep[e.g.,][]{Ho2008,Yuan2014}. To explore the impact of supernova ejecta parameters, we further simulate two different cases, \textit{Model-1} and \textit{Model-2} considering the different kinetic energy and explosion position. The key fiducial parameters are summarized in Table~\ref{tab:parameter}, and detailed equations and complete initial profiles are provided in Appendix~\ref{app:setup}.

We employ the publicly available code \texttt{Athena++} \citep{Stone2020} to perform three-dimensional hydrodynamic simulations of the evolution of SN ejecta near an SMBH. The simulation domain spans $r \in [6 R_{\rm g}, 300 R_{\rm g}]$, $\theta \in [0, \pi/2]$, and $\phi \in [0, 2\pi]$, using a logarithmically uniform grid in the radial direction with $96 \times 96 \times 192$ cells. Outflow boundary conditions are applied in the $r$-direction, polar boundary conditions are applied in the $\theta$-direction, and periodic boundary conditions are applied in the $\phi$-direction. The simulation is evolved until the TDS settles into a stable accretion disk.

\begin{table} [h]
\centering
\caption{Part model parameters in simulations.}
\begin{tabular*}{\linewidth}{c|@{\extracolsep{\fill}}cccc}
\hline
   & \multicolumn{4}{c}{\textbf{SN ejecta}} \\ \hline
    & $M_{\rm ej}$ & $E_{\rm ej}$     & $R_{\rm ej}$ & $v_{\rm ej}$ \\ 
    & ($M_\odot$)  & (erg)            & ($R_{\rm g}$) & (cm/s) \\ \hline
\textit{Model-0}    & 10 & $2\times10^{51}$  & 100 & $3.2\times10^8$ \\ \hline
\textit{Model-1}    & 10 & $1\times10^{52}$  & 100 & $7.1\times10^8$ \\ \hline
\textit{Model-2}    & 10 & $1\times10^{52}$  & 300 & $7.1\times10^8$ \\ \hline
\end{tabular*}
\label{tab:parameter}
\end{table}

\begin{figure*} 
  \centering
  \includegraphics[width=1.0\linewidth]{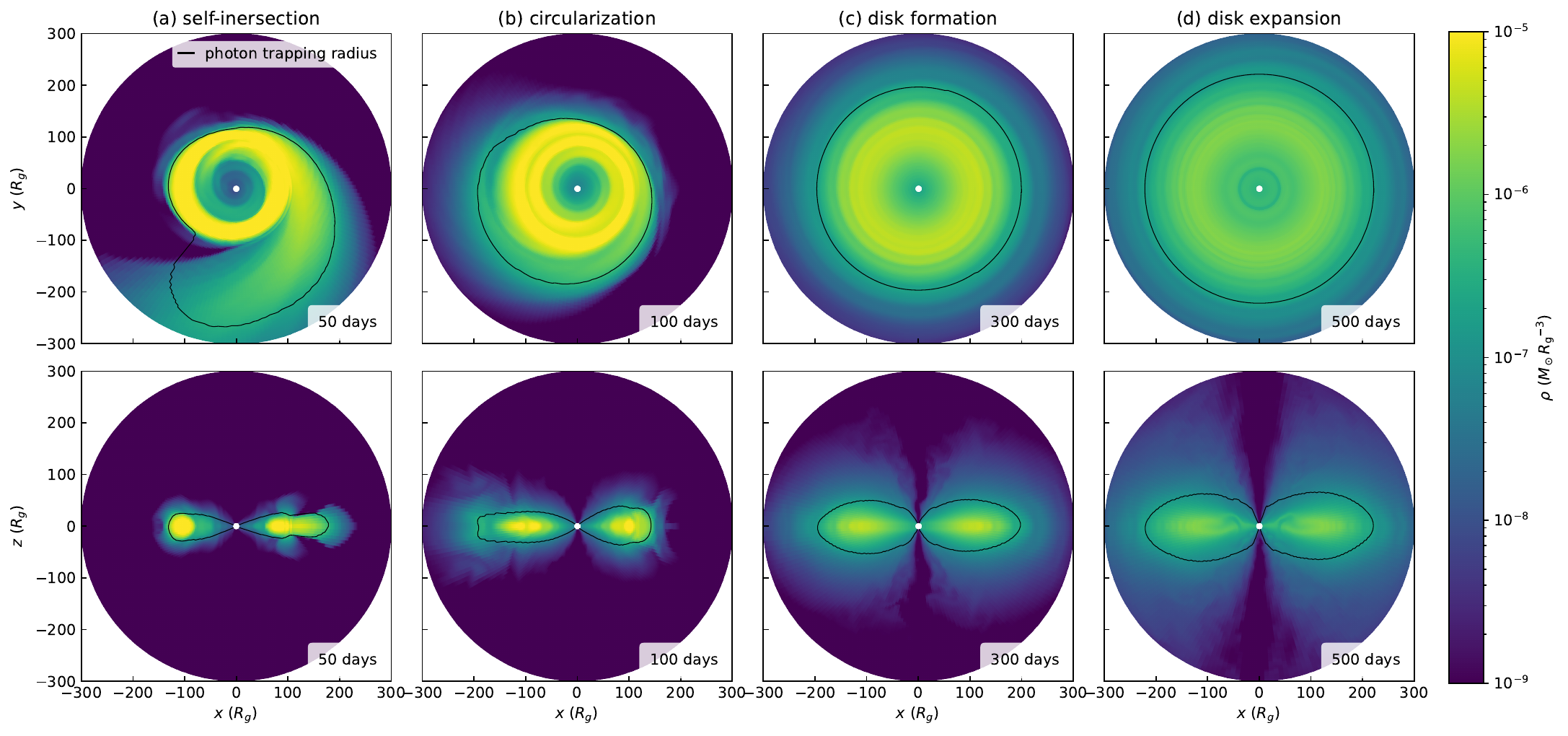}
  \caption{Temporal evolution of the density distribution for the TDS in \textit{Model-0} at several typical snapshots. The time is measured since the SN explosion. The black lines indicate the photon trapping radius \( R_{\rm trap} \). }
  \label{fig:rho_cc}
\end{figure*}

\begin{figure*} [!t]
  \centering
  \includegraphics[width=0.7\linewidth]{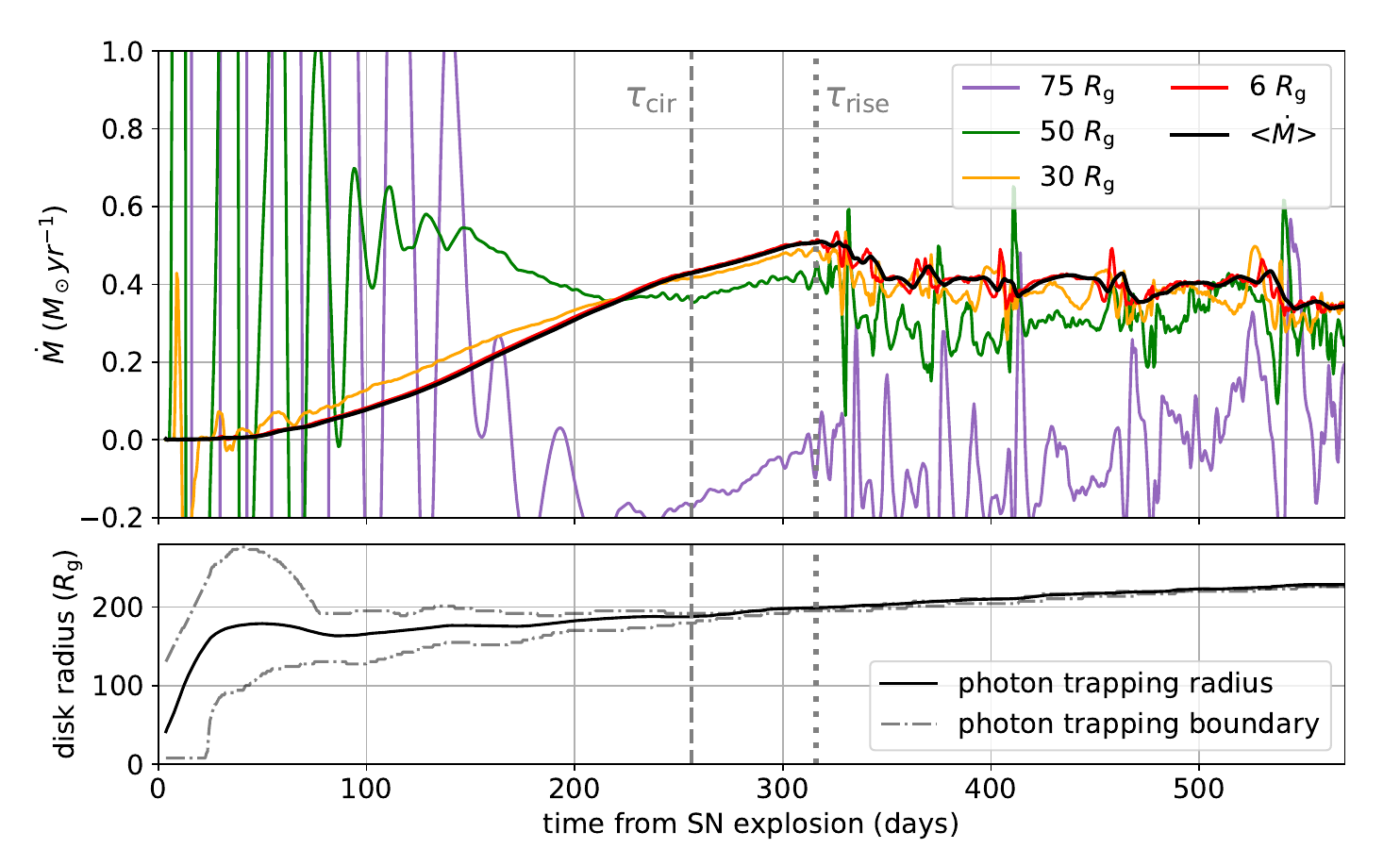}
    \caption{Temporal evolution of the accretion rate (top) and the photon trapping radius (bottom) in \textit{Model-0}. In the top panel, colored solid lines denote the net mass flux at different radii, while the black solid line shows the weekly averaged accretion rate at \(R = 6\,R_{\rm g}\). The vertical dashed line and dotted line mark the circularization timescale \(\tau_{\rm cir}\) and rise timescale \(\tau_{\rm rise}\) (time of peak accretion), respectively. The bottom panel shows the mid-plane photon trapping region equivalent radius (black solid line), with its maximum (\(R_{\rm trap,mid-max}\)) and minimum (\(R_{\rm trap,mid-min}\)) azimuthal boundary indicated by gray dash-dotted lines. We define the circularization timescale \(\tau_{\rm cir}\) as the time after which the condition \(|R_{\rm trap,mid-max}-R_{\rm trap,mid-min}|/R_{\rm trap,mid-max}<5\%\) is satisfied in the mid-plane (vertical gray dashed line).}
  \label{fig:massdot_cc}
\end{figure*}

\section{Results} \label{sec:model}

When a supernova occurs near an SMBH with expansion velocities that are much smaller than the escape velocity, the initially spherical ejecta can be gravitationally captured, forming streams on complex bound orbits. The resulting TDS streams self-intersect near the pericenter, exchange angular momentum, and eventually circularize to form an accretion disk (see Figure~\ref{fig:cartoon}.b). Density snapshots at key epochs from our post-explosion simulation based on the fiducial parameters (\textit{Model-0}) are shown in Figure~\ref{fig:rho_cc}. We present the detailed hydrodynamical and accretion-rate evolution in the following sections.

\begin{figure*} [!t]
  \centering
  \includegraphics[width=1.0\linewidth]{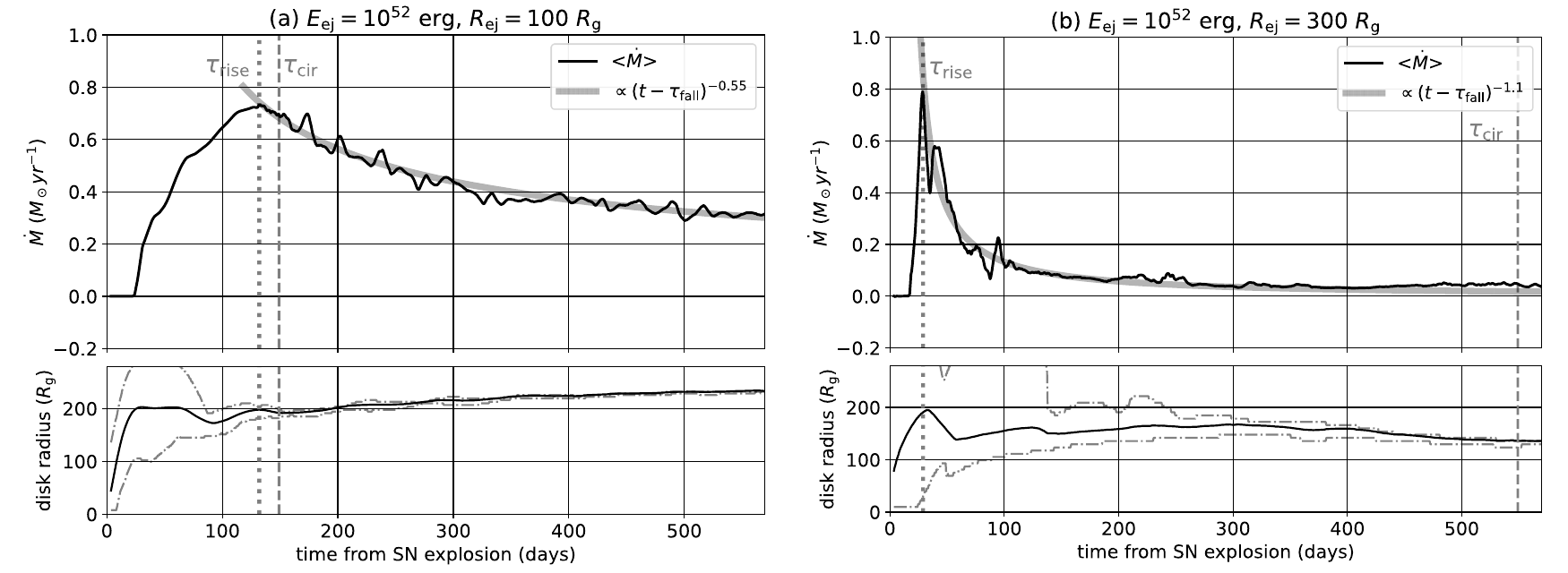}
  \caption{Same as Figure \ref{fig:massdot_cc} but for SN ejecta with \(E_{\rm ej}=10^{52}\) erg, \(R_{\rm ej}=100 R_{\rm g}\) (\textit{Model-1}) and \(R_{\rm ej}=300 R_{\rm g}\) (\textit{Model-2}), respectively. The semi-transparent grey lines indicate power-law declines of the accretion rate.}
  \label{fig:massdot_com}
\end{figure*}

\subsection{Hydrodynamics Process} \label{sec:circularization}

Figure~\ref{fig:rho_cc}a-\ref{fig:rho_cc}c illustrate the self-intersection and circularization stages of the TDS. Initially, the spherical SN ejecta expand beyond the tidal limits, where the tidal forces from the SMBH stretch the ejecta into a diffuse, elongated stream. The low-eccentricity TDS stream undergoes strong self-intersection, which efficiently dissipates angular momentum and drives inward migration. To track the circularization process, Figure~\ref{fig:rho_cc} shows the photon trapping radius, \(R_{\rm trap}\), which is defined as the radius where the photon diffusion timescale \(\tau_{\rm diff} \simeq \tau_o R/c\) equals the electron dynamical timescale \(\tau_{\rm dyn} \simeq R/v\) where \(\tau_o\) is the optical depth and \(v\) is the gas velocity. This condition can be expressed as \(\int_{R_{\rm trap}}^{\infty} \rho(R)\,\kappa_{\rm s}\,dR \simeq c/v_{\rm trap}\), where \(\kappa_{\rm s}\) is electron scattering opacity and \(v_{\rm trap}\) is velocity at \(R_{\rm trap}\) \citep{Rees1978}. The temporal evolution of the \(R_{\rm trap}\) in mid-plane is shown in the bottom panel of Figure~\ref{fig:massdot_cc}. After \(\sim 80\) days, the irregular TDS stream forms precessing ring-like structures within \(\sim 150 R_{\rm g}\) (see Figure~\ref{fig:rho_cc}b). By \(t \sim 256\) days, the stream settles into a stable circular structure (see Figure~\ref{fig:rho_cc}c), indicating the completion of circularization and defined as the timescale \(\tau_{\rm circ}\)..

Figures~\ref{fig:rho_cc}c-\ref{fig:rho_cc}d show the TDS disk structure and its subsequent expansion, which is also reflected in the growing trapping region. After \(t \sim 300\) days, the TDS stream forms a stable disk with an extended trapping region of \(\sim 200 R_{\rm g}\). Most of the material concentrates near the mid-plane, accompanied by outflows at large polar angles. At a later stage, the disk expands beyond the simulation domain, so our measured trapping region may underestimate its true size. Nevertheless, the overall trend of continuous outward expansion is evident.

\subsection{Accretion-Rate Evolution} \label{sec:disk}

To further investigate the disk evolution, Figure~\ref{fig:massdot_cc} shows the net mass flux at different radii. The black solid line shows the weekly averaged net mass flux at $R = 6\,R_{\rm g}$, used to measure the accretion rate onto the SMBH. The accretion rate begins to rise at the fallback timescale $\tau_{\rm fall} \sim 27$ days, estimated from the innermost orbital period $\tau_{\rm fall} \sim 2 \pi \sqrt{{R_{\rm ej}^3}/({GM_{\rm BH}})}\left(1+2\sqrt{{R_{\rm ej}E_{\rm ej}}/({GM_{\rm BH}M_{\rm ej})}}\right)^{-{3}/{2}}$. It then increase and reach a maximum value of $\dot{M} \sim 0.4\,M_\odot\,\mathrm{yr}^{-1} \sim 0.2\,\dot{M}_{\rm Edd}$ at the rise timescale \(\tau_{\rm rise}\sim 316\) days. For adopted fiducial parameters, the accretion phase is expected to last up to \( {M_{\rm ej}}/{\dot{M}}\sim 25\ \mathrm{yr}\), owing to the large SN ejecta mass. Therefore, the late-time evolution in the decay phase is not fully presented in our simulation due to its long timescale and computational constraints.

The accretion-rate evolution of the TDS depends critically on the initial ejecta velocity \(v_{\rm ej}\) and launch position \(R_{\rm ej}\). For comparison, we present two additional cases in Figure~\ref{fig:massdot_com} with higher SN energy \(E_{\rm ej}=10^{52}\) erg launched at \(R_{\rm ej}=100\,R_{\rm g}\) (\textit{Model-1}) and \(R_{\rm ej}=300\,R_{\rm g}\) (\textit{Model-2}). In \textit{Model-1}, the accretion rate exhibits a slow power-law decline, \(\dot{M} \propto (t-\tau_{\rm fall})^{-0.55}\), after reaching a peak of \(\dot{M} \sim 0.7\,M_\odot\,\mathrm{yr}^{-1}\) at \(t \sim 132\) days. The bottom panels show that circularization completes at \(\tau_{\rm cir} \sim 149\) days. In this case, \(\tau_{\rm rise} < \tau_{\rm cir}\), indicating that angular momentum redistribution in the accretion flow is relatively inefficient. In \textit{Model-2}, the accretion rate reaches its maximum of \(\dot{M} \sim 0.8\,M_\odot\,\mathrm{yr}^{-1}\) at \(\tau_{\rm rise} \sim 29\) days. Most of the outer ejecta material is accreted within \(\sim 400\) days before full circularization, where \(\tau_{\rm rise} \ll \tau_{\rm cir}\) and the accretion rate declines more steeply as \(\dot{M} \propto (t-\tau_{\rm fall})^{-1.1}\).

\section{Summary and Discussion} \label{sec:discussion}

We carry out three-dimensional hydrodynamic simulations of the tidal distortion of a supernova that explodes close to an SMBH. In our simulations, a fraction of the ejecta is captured by the gravitational potential, undergoes strong self-intersection, and eventually forms an accretion disk. The fiducial simulation yields a steady accretion rate \(\dot{M}\sim0.4\,M_\odot\,\mathrm{yr^{-1}}\), which corresponds to a luminosity of \(L_{\rm TDS}=\eta\dot{M}c^2\sim2\times10^{45}\,\mathrm{erg\,s^{-1}}\) with radiative efficiency \(\eta\sim0.1\). Two generic accretion behaviors are identified: (1) a quasi-steady plateau phase when circularization is efficient and a new disk is established, and (2) a power-law decline when fallback and dissipation proceed more gradually. These results demonstrate that TDS can produce TDE-like accretion episodes and peak energetics comparable to those of observed luminous nuclear transients. We note that our present calculations omit radiation transport, magnetic fields, and full general-relativistic effects; therefore, quantitative predictions for light curves and spectra require radiation-GR(M)HD follow-up studies.

In Section~\ref{sec:model}, we show that the accretion properties of the TDS are primarily governed by the circularization timescale \(\tau_{\rm cir}\) and the rise timescale \(\tau_{\rm rise}\), which determine whether the accretion rate evolves into a plateau-like phase resembling a stable disk or follows a power-law decline similar to the initial fallback rate. In our fiducial simulation (\textit{Model-0}), the timescale to reach the peak accretion rate is \(\tau_{\rm rise} \sim 5.8\,R_{\rm ej}/v_{\rm ej}\), a scaling that also holds for \textit{Model-1} where SN velocities are much smaller than the escape velocity. In \textit{Model-2}, the rise timescale approaches the free-fall timescale, indicating that the outer SN ejecta undergoes nearly radial infall with minimal self-intersection, due to an ejection velocity approaching the escape velocity. The circularization timescale can be expressed as \(\tau_{\rm cir} \sim f_{\rm cir}\,\tau_{\rm fall}\) like that in TDEs \citep{Bonnerot2017,Chen2021}, where the factor \(f_{\rm cir}\) measures the circularization efficiency. Based on our simulations, we find \(f_{\rm cir} \sim 9.4\), 7.0 and 6.2 for \textit{Model-0,-1,-2} respectively, where the stronger self-intersection enhances angular momentum dissipation with increasing values of \(v_{\rm ej}/v_{\rm K}\). Higher explosion velocities at a larger launch site result in a steeper power-law evolution of the accretion rate (\(\tau_{\rm rise} \ll \tau_{\rm cir}\)), due to incomplete redistribution of energy and angular momentum during self-intersection. We also note that the initial ejecta size can slightly affect the accretion-rate evolution, which requires higher-resolution simulations for further investigation.

Although our simulations are conducted for a black hole mass of \(M_{\rm BH} = 10^8\,M_\odot\), the characteristic timescales scale as \(GM_{\rm BH}/c^3\). Consequently, TDS disks around smaller black holes (\(M_{\rm BH} \lesssim 10^{7.5}\,M_\odot\)) can sustain super-Eddington accretion rates over correspondingly shorter durations. Moreover, the accretion rate also scales with the ejecta mass, \(\dot{M} \propto M_{\rm ej}\). Very massive supernovae, such as pair-instability supernovae (PISNe) with \(M_{\rm ej} \sim 100\,M_\odot\) \citep{Fryer2001,Gal-Yam2009}, can therefore produce long-lived, super-Eddington disks. In this super-Eddington regime, a photosphere may form around the TDS disk, powered by radiation-pressure-driven winds or self-intersection-induced outflows, analogous to TDEs \citep[e.g.,][]{Dai2018,Curd2021,Qiao2025}. Consequently, TDSs are expected to exhibit multiwavelength signatures similar to TDEs, but they are predicted to be more luminous, exhibit smoother light curves, and persist for longer durations because of the larger progenitor mass and more efficient angular momentum redistribution. Unlike TDEs, the TDSs can naturally occur in galaxies hosting SMBHs of any mass and can produce extreme flares even for \(M_{\rm BH} \gtrsim 10^8\,M_\odot\). Spectroscopically, the heavy-element-enriched SN ejecta should imprint strong metallic features in TDS, which may lead to some strong metal lines (e.g., O, Si, Fe) that can be tested in future spectroscopic observations. As shown above, we expect that the TDSs are preferentially appear in low-luminosity AGNs. It should be noted that the dynamics and stellar evolution may be significantly affected by dense environments, which have begun to attract attention recently \citep[e.g.][]{Wang2021,Cantiello2021,Dittmann2021}. For example, dense disk environments promote rapid stellar accretion and strong envelope stripping, leading to compact massive stars with radii of only a few $R_\odot$ in AGN disks \citep[e.g.,][]{Cantiello2021,Jermyn2021}. The disruption radius of these compact stars is smaller than that of typical main sequence stars with similar mass. They are expected to evolve and easily explode as supernovae after they stop accretion in low density ADAF. Further investigations of stellar processes in dense environment and of AGN evolution are crucial for estimating the TDS event rate, and we hope to address these issues in future work.  

Several extreme, TDE-like flares associated with overmassive SMBHs have been reported (e.g., ASASSN-15lh in \citealt{Dong2016,Leloudas2016}; J2245+3743 in \citealt{Graham2025}; and three ENTs in \citealt{Hinkle2025}). These transients are exceptionally luminous ($L\sim10^{45}\!-\!10^{46}\ \mathrm{erg\ s^{-1}}$), long-lived (hundreds to thousands of days), and occur in host galaxies with SMBH masses $M_{\rm BH}\gtrsim10^{8.5}\,M_\odot$. The host galaxies of these transients commonly exhibit weak AGN activity prior to the strong flares, with bolometric Eddington ratios less than a few percent\citep{Kruhler2018,Subrayan2023,Hinkle2025,Graham2025}. Importantly, unlike TDEs with typical decline as $\propto t^{-5/3}$ (and even steeper for more massive stars; \citealt{Guillochon2013}), these events display much shallower post-peak declines, approximately $\propto t^{-2/3}$ \citep[see][for more details]{Graham2025}. Both the environments and the light curves suggest that the physical mechanism driving these transients may differs from that of typical TDEs. Our TDS model in LLAGNs provides an explanation for these ENTs/TDE-like flares. 

Although the physical mechanism driving CLAGN remains uncertain, variations of SMBH accretion rate have been widely discussed as a plausible explanation \citep[e.g.,][]{Eracleous1995,Elitzur2014,Merloni2015,Wang2018,Wang2024,Wang2024b,Li2025}. Observationally, turn-on CLAGN events preferentially occur in low-activity AGN with Eddington ratios of only a few percent \citep[e.g.,][]{Noda2018,Lyu2021,Lyu2022,Liu2022,Hon2022,WangS2024,Guo2025,Qiao2025}. In such an AGN disk, a supernova explosion can inject a substantial amount of mass into the accretion flow, thereby enhancing the accretion rate and potentially triggering a trun-on CLAGN that can persist for years (e.g., ASASSN-17jz in \citealt{Holoien2022}; 1ES 1927+654 in \citealt{Trakhtenbrot2019,Li2024}). The TDS mechanism, by supplying a large significant amount of material in LLAGN environments, therefore, offer an explanation for turn-on CLAGN behaviour.

The evolution of a TDS begins with the supernova explosion, and our model will lead to an SN flare before the accretion-powered flare. But the typical core-collapse SNe peak at $L_{\rm peak}\sim10^{42}\ \mathrm{erg\ s^{-1}}$ \citep[e.g.,][]{Li2011,Richardson2014}, which is much fainter than both the subsequent TDS flare and the host nuclear luminosities of several reported ENTs (typical $L_{\rm AGN}\sim10^{45}\ \mathrm{erg\ s^{-1}}$; e.g., \citealt{Hinkle2025,Graham2025}), and is therefore difficult to detect against bright nuclear backgrounds. We anticipate that future detections of unambiguous precursors would provide valuable confirmation of the TDS scenario.

\bibliography{cite}{}

\bibliographystyle{aasjournal}

\begin{appendix}

\section{Numerical Setup} \label{app:setup}

For the initial disk background, we adopt the hydrostatic‐equilibrium disk model of \citet{Nelson2013}, in which the mid-plane temperature and density follow power laws,
\begin{equation}
    T(R) = T_0 (\frac{R}{R_{\rm g}})^{-q},
    \label{eq:T}
\end{equation}
\begin{equation}
    \rho(R,Z) = \rho_0 (\frac{R}{R_{\rm g}})^{-p}\exp\left[\frac{GM_{\rm BH}}{c_s^2}\left(\frac{1}{\sqrt{R^2+Z^2}}-\frac{1}{R}\right)\right],
    \label{eq:rho}
\end{equation}
\begin{equation}
    \Omega(R,Z) = \Omega_{\rm K}\left[\frac{qR}{\sqrt{R^2+Z^2}}+(1-q)-(p+q)\left(\frac{H}{R}\right)^2\right]^{1/2},
    \label{eq:vk}
\end{equation}
where \(\Omega_{\rm K}=\sqrt{GM_{\rm BH}/R^3}\) with \(R\) is the distance from the central SMBH, and the vertical scale height is \(H=c_{\rm s}/\Omega_{\rm K}\). Adopting the self-similar ADAF solution of \citet{Narayan1994} with \(q=1\) and \(p=1.5\), we obtain characteristic mid-plane values of \(T_0 \sim 10^{12}\,\mathrm{K}\) and \(\rho_0 \sim 10^{-13}\,\mathrm{g\,cm^{-3}}\), with a dimensionless shear viscosity parameter \(\alpha=\mu/(\rho c_s H)=0.2\).

And we adopt an isotropic SN ejecta model with an exponential density profile, as described in the simulations of \citet{Grishin2021}. Such profiles are commonly employed in SN ejecta models \citep[e.g.,][]{Nomoto1984,Matzner1999}, and follow
\begin{equation}
    v_\text{ej,ini}(r) = v_\text{ej} \left(\frac{r}{r_{\rm ej} }\right),
    \label{eq:v_sn}
\end{equation}  
\begin{equation}
    \rho_\text{ej,ini}(r) = \sqrt{6}\,\rho_\text{ej}\,\exp\!\Bigl(-\sqrt{6}\frac{r}{r_{\rm ej}}\Bigr),
    \label{eq:rho_sn}
\end{equation}  
where \(r\) is the distance from the center of SN ejecta, \( r_{\rm ej} =  v_{\rm ej}t_0\) is the characteristic size with the simulation start time since the SN explosion \(t_0\), and \( \rho_{\rm ej} =  M_{\rm ej}/(4\pi r_{\rm ej}^3/3) \) density. The total energy of SN is $E_{\rm ej} = E_{\rm ej, kinetic} + E_{\rm ej, thermal}$. We adopt the equipartition assumption \(E_{\rm ej, kinetic} = E_{\rm ej, thermal}\), with a uniform temperature inside \(r_{\rm ej}\) and one tenth of the central temperature in the outer shell, commonly used in analytic ejecta models. For feasibility, we adopt an initial large ejecta size of \( r_{\rm ej} \sim 7.5 R_{\rm g} \) and cut off the profile at \( r = 4 r_{\rm ej} \). Although this overestimates the spherical ejecta size, the effect on global dynamics is negligible.

We employ the publicly available code \texttt{Athena++} \cite{Stone2020} to solve the time-dependent equations of hydrodynamics with a viscosity term, 
\begin{equation}
\frac{\partial \rho}{\partial t}
+ \nabla \cdot (\rho \mathbf{v})
= 0,
\end{equation}
\begin{equation}
\frac{\partial (\rho \mathbf{v})}{\partial t}
+ \nabla \cdot (\rho \mathbf{v}\otimes\mathbf{v})
= -\,\nabla P
- \rho\,\nabla\Phi
+ \nabla\cdot\mathbf{T},
\end{equation}
\begin{equation}
\frac{\partial E}{\partial t}
+ \nabla\cdot\bigl[(E + P)\,\mathbf{v}-\mathbf{T}\cdot\mathbf{v}\bigr]
= -\,\rho\,\mathbf{v}\cdot\nabla\Phi.
\end{equation}
The total gas energy density is given by \(E = P/(\gamma - 1) + \frac{1}{2} \rho \mathbf{v}^2\) and gas temperature is \(T=P\mathcal{M}/\mathcal{R}\rho\), where the adiabatic index is \(\gamma = 5/3\), \(\mathcal{R}\) is the ideal gas constant and the mean molecular weight \(\mathcal{M}=0.6\). The \(\mathbf{T}\) is a viscous stress tensor with the dynamical viscosity parameter \(\mu=\rho \alpha c_{\rm s}H\). The gravitational potential of the central black hole is included as \(\Phi(r) = -{GM_{\rm BH}}/{r}\) outside the inner boundary \(6R_{\rm g}\).

\end{appendix}

\end{document}